\documentclass[12pt]{article}
\usepackage{epsfig}

\usepackage{amssymb}
\usepackage{amsfonts}

\usepackage{graphicx}

\newtheorem{lem}{Lemma}

\newtheorem{prop}[lem]{Proposition}
\newtheorem{thm}{Theorem}

\newtheorem{pozn}{Remark}

\def\bp{{\em Proof.} }
\def\ep{$\Box $}

\def\be{\begin{equation}}
\def\ee{\end{equation}}
\def\ba{\begin{array}{c}}
\def\ea{\end{array}}

\def\ben{\[}
\def\een{\]}

\newcommand{\bea}{\begin{eqnarray}}
\newcommand{\eea}{\end{eqnarray}}

\newcommand{\bbr}{\br\!\br}
\newcommand{\kkt}{\kt\!\kt}

\newcommand{\kt}{\rangle}
\newcommand{\br}{\langle}
\begin{document}

\titlepage

\vspace{.35cm}

 \begin{center}{\Large \bf

Solvable quantum lattices with nonlocal non-Hermitian endpoint
interactions

  }\end{center}

\vspace{10mm}

 \begin{center}

 {\bf Miloslav Znojil}

 \vspace{3mm}
Nuclear Physics Institute ASCR,

250 68 \v{R}e\v{z}, Czech Republic

{e-mail: znojil@ujf.cas.cz}

\vspace{3mm}

%

\end{center}

\vspace{5mm}


\section*{Abstract}

Discrete multiparametric 1D quantum well with ${\cal PT}-$symmetric
long-range boundary conditions is proposed and studied. As a
nonlocal descendant of the square well families endowed with Dirac
(i.e., Hermitian) and with complex Robin (i.e., non-Hermitian but
still local) boundary conditions, the model is shown characterized
by the survival of solvability in combination with an enhanced
spectral-design flexibility. The solvability incorporates also the
feasibility of closed-form constructions of the physical
Hilbert-space inner products rendering the time-evolution unitary.

\subsection*{KEYWORDS}
.

 exactly solvable quantum models;

non-Hermitian boundary conditions;

new nonlocal boundary conditions;

physical inner products;

\newpage

\section{Introduction\label{alfa} }

An active interest of researchers in quantum lattices {\it alias}
quantum chain models {\it alias} tridiagonal-matrix Hamiltonians
 \be
 H_{chain}=
 \left[
 \begin {array}{ccccc}
  a_0&b_0&0 &\ldots&0
  \\
 c_1&a_1&b_1&\ddots&\vdots
 \\
 0&\ddots&\ddots&\ddots &0
 \\
 \vdots&\ddots&c_{N-1}&a_{N-1}&b_{N-1}
 \\
 0&\ldots&0&c_N&a_N
 \end {array} \right]\,
 \label{LuKa8}
 \ee
dates back to the pioneering applications of these Hamiltonians in
organic chemistry in the thirties \cite{Hukel}. Still, their study
belongs to the mainstream activities, say, in the tight-binding
descriptions of electronic structures in solids \cite{Goringe}, etc.
The models of this type also helped to clarify some non-variational
features of Green's functions in particle physics \cite{Varma} and
they are currently serving as an exemplification of several
gain-and-loss-related phenomena in optics \cite{Longhi}. Last but
not least, the use of models (\ref{LuKa8}) threw new light on some
questions of the emergence and confluence of the Kato's exceptional
points (KEP, \cite{Kato}) in perturbation theory
\cite{maxi,maximal}.

In our recent paper \cite{I} we decided to pay a detailed attention
to one of the simplest quantum models of this type, characterized by
the next-to-trivial $(N+1)$ by $(N+1)$ matrix Hamiltonian
 \be
 H^{(N+1)}(z)=
 \left[
 \begin {array}{cccccc}
  2{}- z&-1&0 &\ldots&0&0
  \\
 -1&2{}&-1&0&\ldots&0
 \\
 {}0&-1&2&\ddots &\ddots&\vdots
 \\
 \vdots&\ddots&\ddots&\ddots &-1&0
 \\
 {}0&\ldots&0&-1&2{}&-1
 \\
 {}0&0&\ldots&0&-1&2{}- z^*
 \end {array} \right]\,
 \label{uKa8}
 \ee
with a single complex non-real parameter $z \in \mathbb{C} \setminus
\mathbb{R}$. We revealed that in spite of its manifest
non-Hermiticity the model may be perceived as belonging to the
conventional quantum theory in which it generates a unitary
evolution of the system in question.

The results of paper \cite{I} offered an explicit example and
illustration of certain recent, not entirely conventional
implementations of the abstract principles of quantum theory. In
essence, these implementations are based on an innovative
parity-times-time-reversal-symmetric (${\cal PT}-$symmetric, PTS,
\cite{Carl}) presentation {\it alias} pseudo-Hermitian
representation (PHR, \cite{ali}) {\it alias} three-Hilbert-space
(THS, \cite{Geyer,SIGMA}) form of the formalism of quantum theory
(see Appendix A for a compact summary of these ideas).

In paper \cite{I} we just confirmed that in practice, the costs of
the PTS/PHR/THS enhancement of the flexibility of models may be
reasonable and acceptable. Our main result was that we managed to
fulfill the technically most difficult task of the PTS/PHR/THS
theory and reconstructed the nontrivial ``standard'' representation
${\cal H}^{(S)}$ of the physical Hilbert space in which the
manifestly non-Hermitian Hamiltonian (\ref{uKa8}) becomes Hermitized
so that, in other words, the evolution in time acquires the standard
unitary interpretation.

A formal simplicity of the discrete model (\ref{uKa8})
may be perceived as inherited
from its differential-operator predecessor of Ref.~\cite{sdavidem}.
In parallel,
it should be emphasized that these two models also share
the
physical motivation and an immediate phenomenological appeal
which was verbalized in Ref.~\cite{Coronado}
and which appeared to lie in the existence of a
close relationship between the bound-state and
scattering experimental data (cf. also Refs.~\cite{Hernandez,data} in this context).

All these observations provoked, naturally, a
search for the generalizations which would eventually go beyond
the tridiagonal matrix structure (\ref{LuKa8}) of the model.
A few new results
obtained in this direction will be presented in what follows. First
of all, our inspiration by PTS matrix~(\ref{uKa8}) will lead, in a
way described in section \ref{sedv}, to its partitioning and to its
subsequent multiparametric PTS generalization. In section
\ref{sedvbe} we shall illustrate some descriptive merits of spectra
provided by this generalization. The survival of solvability of the
model will finally be demonstrated, constructively, in section
\ref{solgiga}.

Due to the non-Hermiticity of the Hamiltonian, we will have to
address also the above-mentioned problem of construction of the
correct physical representation Hilbert space. This will be done in
several sections. Firstly, the general recipe will be presented in
section \ref{secty}. Two alternative methods of construction of the
correct inner-product metric $\Theta$ will be described. We shall
point out that both parts of the construction of the bound state
solutions (viz., the construction of the wave functions and of the
metric) appear closely interconnected in a way which may be
perceived as an explanation why, in spite of its enhanced
flexibility, our generalized model remains tractable
non-numerically.

An alternative approach capable of producing some of the metrics in
a friendlier sparse-matrix form will be then conjectured and studied
in sections \ref{sepet} and  \ref{sepetbe}. In the former section we
shall put more emphasis upon the phenomenological aspects of the
metrics reflected by specific choices of optional parameters. In the
latter section we point out that the advantages of the sparse
metrics are accompanied by some obstacles. They will be exemplified
by the emergence of certain critical parametric hypersurfaces of the
loss of positivity of the metric, i.e., in the language of physics,
of a phase-transition loss of the usual probabilistic interpretation
of the model in question.

In section \ref{pst} we then add a brief remark on what happens,
beyond the realm of quantum theory, when the invertibility survives
but the positivity of the metric $\Theta$ is lost. Purely formally,
matrix $\Theta$ then acquires a new meaning of pseudometric
$\tilde{\cal P}$. Although it still renders the given Hamiltonian
self-adjoint in an {\it ad hoc} Krein or Pontriagin space
\cite{Langer}, its physical meaning can only survive, say, in some
less ambitious, non-unitary effective-theory implementations
\cite{Jones,Cannata}.

The last section \ref{summary} will offer a brief summary of our
results. In order to keep our present paper sufficiently
self-contained we decided to complement the main text by two
Appendices. In the first one we recall Ref.~\cite{SIGMA} and
summarize the key features of the PTS/PHR/THS quantum mechanics as a
formalism which is based on the simultaneous use of a {\em triplet}
of alternative representation Hilbert spaces $\{ {\cal H}^{(P)},
{\cal H}^{(F)}, {\cal H}^{(S)} \}$. In Appendix B we shall turn
attention to the well known (though rarely emphasized) specific fact
that in an experimental setting, {\em any} generic PTS/PHR/THS model
controlled by a non-Hermitian Hamiltonian is in fact nonlocal. This
observation may be read as an idea which, initially, motivated our
present considerations.

\section{The model\label{sedv}}

\subsection{Local boundary conditions\label{sedvuh}}

In virtually any textbook on conventional quantum mechanics one
finds the formalism illustrated by a one-dimensional square well. In
such a model the wave functions obey differential Schr\"{o}dinger
equation
 \be
 -\frac{d^2}{dx^2}\psi_n(x)=E_n\,\psi_n(x)
 \label{SE}
 \ee
and the Dirichlet boundary conditions are imposed at a pair of
boundary points,
 \be
 \psi_n(\pm L)=0\,,
 \ \ \ \ \ \ \  L \in \mathbb{R} \setminus \{0\}\,.
 \label{Dirichlet}
 \ee
In \cite{sdavidem}, similarly, the PTS/PHR/THS formalism was
illustrated using a replacement of Eq.~(\ref{Dirichlet}) by complex
Robin boundary conditions
 \be
  \psi(\pm L)= \frac{\rm i }{\alpha \mp {\rm i}\beta}
  \,\frac{d}{dx}\psi(\pm L)\,,
  \ \ \ \ \ \ \ \
 \alpha \,,\, \beta\, \in \,\mathbb{R}
 \,.
   \label{spojhambcdavid}
 \ee
In this context our subsequent paper \cite{I}
merely replaced {differential} Eq.~(\ref{SE}) by its
{difference}-equation alternative
 \be
 -\phi^{(n)}_{k-1}+2\,\phi^{(n)}_{k}-\phi^{(n)}_{k+1}=
 E^{(n)}\phi^{(n)}_{k}\,,
 \ \ \ \ \ k = \ldots, -1,0,1,\ldots
 \,.
 \label{DSE}
 \ee
We emphasized there that one of benefits of the discretization
(\ref{SE}) $\to$ (\ref{DSE}) lies in a survival of simplicity of
boundary conditions in which two Eqs.~(\ref{spojhambcdavid}) were
merely replaced by their two equally elementary discrete analogues
 \be
 \phi^{(n)}_{-1}=z\,\phi^{(n)}_{0}\,,
 \ \ \ \ \
 \phi^{(n)}_{N+1}=z^*\,\phi^{(n)}_{N}\,,
 \ \ \ \ \ \ \ \ \ \
 \label{RBC}
 \ee
 \ben
 \ \ \ \ \ \ \ \ \ \
 \ \ \ \ \ \ \ \ \ \
 \ \ \ \ \
 z =z(\xi,\zeta)=(1-\xi-{\rm i} \zeta)^{-1} \in \mathbb{C}\,,
 \ \ \ \ \xi, \zeta \in \mathbb{R}\,.
 \een
The role of the original continuous real coordinate $x \in
\mathbb{R}$ was smoothly transferred to its discrete real
counterpart $k \in \mathbb{Z}$. As a consequence the Hamiltonian
itself degenerated to the above-mentioned finite-dimensional matrix
Hamiltonian (\ref{uKa8}).

\subsection{Long-range boundary conditions\label{sedvuhwa}}

Hamiltonian (\ref{uKa8}) has the conventional form of  superposition
$H=T+V$. In contrast to its standard kinetic-energy part $T$, the
form of the local (i.e., diagonal) potential-energy matrix $V$ is
much less standard. Not only that it represents just a
point-supported interaction (i.e., it vanishes everywhere except the
end-points of the discrete-coordinate lattice) but it is also
manifestly non-Hermitian, $V\neq V^\dagger$.

In papers \cite{Jones,Batal} it has been shown that in the ultimate
experimental setup (i.e., in the primary physical Hilbert space
${\cal H}^{(P)}$) {\em both} the locality and the point-interaction
nature get inadvertently lost for a generic non-Hermitian
interaction $V$, local or nonlocal. (in detail the reasons are
recalled in Appendix B below). In other words, the decisive reason
for the preferred choice of matrices $V$ shouldn't be seen in their
diagonality (i.e., in their locality without physical meaning - cf.
comments in \cite{Jones}) but merely in their sparsity. Guaranteeing
that with the growth of the matrix dimension $(N+1)$, the number
$M(V)$ of non-vanishing matrix elements becomes negligible in
comparison with the total number $(N+1)^2$ of all matrix elements.

Naturally, the latter condition is more than perfectly satisfied by
the $M(V)=2$ Hamiltonian (\ref{uKa8}) of paper \cite{I}. We believe,
nevertheless, that the  restriction $M(V)=2$ may be weakened at a
reasonable cost. In our present sequel of the latter study we intend
to generalize boundary conditions to a nonlocal constraint,
therefore. As long as we still intend to keep contact with their
local predecessors, we shall postulate that if $V_{ij} \neq 0$ then
either $i$ or $j$ is equal to $0$ or $N$.

The latter postulate admits all complex interaction matrices
of the following partitioned form
 \be
 V^{(N+1)}(\vec{a},\vec{b},\vec{c},\vec{d})
 =\left[ \begin {array}{c|cccc|c}
  b_0&a_1^*&a_2^* &\ldots&a_{N-1}^*&d_0
 \\
 \hline
 b_1&0&0&\ldots&0&d_1
 \\
 b_2&0&0&\ldots&0&d_2
 \\
 \vdots&\vdots&\vdots&\ddots&\vdots
 &\vdots
 \\
 b_{N-1}&0&0&\ldots&0&d_{N-1}\\
 \hline
 b_N&c_1^*&c_2^*&\ldots&c_{N-1}^*&d_N
 \end {array} \right]\,.
 \label{VeKa8}
 \ee
%
Obviously, this yields a class of multiparametric sparse-matrix
models with $M(V)={\cal O}(N)$. In parallel, in a less formal and
more phenomenologically oriented perspective a truly remarkable
feature of such forces may be seen in their surface-related support
leaving the internal, bulk part of our discrete lattice
uninfluenced. In this sense, models (\ref{VeKa8}) may be perceived
as a certain missing link between the strictly local and some more
general and realistic, spatially smeared thick-boundary-layer
interactions.

\subsection{${\cal PT}-$symmetry constraint}

In conventional quantum mechanics one would further restrict the set
of parameters in Eq.~(\ref{VeKa8}) by the Hermiticity constraint
$V=V^\dagger$. This means that with $a_0^*\equiv b_0$, $c_0^* \equiv
b_N$, $a_N^*\equiv d_0$ and  $c_N^*\equiv d_N$ one would impose the
two additional $(N+1)-$plets of requirements $\vec{a}=\vec{b}$ and
$\vec{c}=\vec{d}$. In our present paper we shall proceed in
different, PTS/PHR/THS spirit. In the light of multiple related
technical challenges we shall, first of all, try to support the
reality of the spectrum of $H \neq H^\dagger$ by an additional,
auxiliary PTS condition $ H{\cal PT}={\cal PT}H$ {\it alias}
parity-pseudo-Hermiticity constraint $ H^\dagger{\cal P}={\cal P}H$.

Using the antidiagonal parity-representing matrix
 \be
 {\cal P}= {\cal P}^{(N+1)}=
  \left[ \begin {array}{ccccc}
 0&0&\ldots&0&1
 \\{}0&\ldots&0&1&0\\
 {}\vdots&
 {\large \bf _. } \cdot {\large \bf ^{^.}}&
 {\large \bf _. } \cdot {\large \bf ^{^.}}
 &
 {\large \bf _. } \cdot {\large \bf ^{^.}}&\vdots
 \\{}0&1&0&\ldots&0
 \\{}1&0&\ldots&0&0
 \end {array} \right]\,
 \label{anago}
 \ee
this leads to the  most general PTS Hamiltonian
 \be
 H^{(N+1)}
 =\left[ \begin {array}{c|ccccc|c}
  2{}-z&{\beta}_{N-1}^*-1&\beta^*_{N-2}&\ldots&\beta_2^*&\beta_1^*&b
 \\
 \hline
 {\alpha}_1-1&2{}&-1&0&\ldots&0&\beta_1
 \\
 \alpha_2&-1&2{}&\ddots&\ddots& \vdots&\beta_2 \\
 \alpha_3&0&-1&\ddots&-1&0&\vdots \\
 \vdots&\vdots&\ddots&\ddots&2{}&-1&\beta_{N-2}\\
 \alpha_{N-1}&0&\ldots&0&-1&2{}&{\beta}_{N-1}-1\\
 \hline
 a&\alpha^*_{N-1}&\alpha^*_{N-2}&\ldots&\alpha^*_2&
 {\alpha}^*_1-1&2{}-z^*
  \end {array} \right]\,.
 \label{HaKa7}
 \ee
In the notation of Eq.~(\ref{VeKa8}) this matrix contains the real
elements $b_N=a$ and $d_0=b$ and the mutually conjugate complex pair
of $b_0=-z$ and $d_N=-z^*$. The new symbols $\vec{\alpha}$ and
$\vec{\beta}$ now denote the $(N-1)-$dimensional complex vectors
such that the $M(V)=2$ non-Hermitian model (\ref{uKa8}) of paper
\cite{I} re-emerges here as the simplest special case with  $a=b=0$
and $\vec{\alpha}=\vec{0}$ and $\vec{\beta}=\vec{0}$.

\section{Spectral-design applications\label{sedvbe}}

In contrast to the matrix-boundary models (\ref{VeKa8}) which are
chosen Hermitian and which would generate the robustly real spectra,
the violation of the Hermiticity may render the parametric
dependence of the spectrum much more versatile. Typically, for our
present ${\cal PT}-$symmetric model (\ref{HaKa7}) the reality of the
levels may be lost at the KEP values of parameters. Thus, in
principle, the model may be expected to offer a wealth of unusual
spectral patterns. Also the multiparametric nature of our
matrix-boundary conditions would enhance the flexibility of any
phenomenologically-oriented spectral design.

\subsection{$N=4$ example}

\begin{figure}[h]                     
\begin{center}                         
\epsfig{file=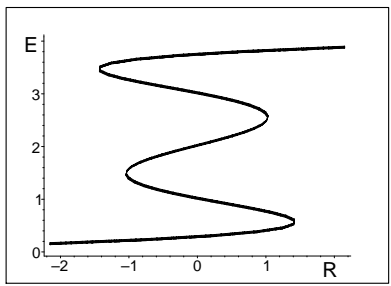,angle=270,width=0.3\textwidth}
\end{center}                         
\vspace{-2mm}\caption{The $R-$dependence of energies in potential
(\ref{tripl}) at $\varrho=\varrho_a$.
 \label{fi05}}
\end{figure}

A sufficiently representative sample of the spectral patterns may be
obtained even via one-parametric variations of our matrix boundaries
(or, better, row-and-column, vectorial boundaries). Some of these
patterns may be even revealed at the smallest matrix dimensions
$N+1$. We found it instructive to use $N+1=5$ and to construct
Hamiltonians $ H=H(R)= T+R\,V^{(\varrho)} \,$ with a single real
variable coupling $R$, with a discrete Laplacean for the kinetic
energy (in the notation of Eq.~(\ref{uKa8}) we have
$T=H^{(N+1)}(0)$) and with potential
  \be
 V^{(\varrho)}=\left[ \begin {array}{ccccc}
 \sigma_4-i\,\tau_4&\sigma_3-i\,\tau_3&\sigma_2-i\,
 \tau_2&\sigma_1-i\,\tau_1&\sigma_0
 \\
 \noalign{\medskip}0&0&0&0&\sigma_1+i\,\tau_1
 \\
 \noalign{\medskip}0&0&0&0&\sigma_2+i\,\tau_2
 \\
 \noalign{\medskip}0&0&0&0&\sigma_3+i\,\tau_3
  \\
  \noalign{\medskip}0&0&0&0&\sigma_4+i\,\tau_4
  \end {array} \right]
 \label{tripl}
 \ee
where $\sigma_m=$ $0$ or $1$ and $\tau_n=$ $0$ or $1$. In this way
the model is defined in terms of the two quintuplets of  binary
digits forming a complex index
 $$\varrho= (\sigma_4,\sigma_3,\sigma_2,\sigma_2,\sigma_0)
 +i\,(\tau_4,\tau_3,\tau_2,\tau_1,\tau_0).
 $$
Let us now pick up one of the simplest
indices
 $$\varrho_a= (0,0,0,0,1) +i\,(0,0,0,0,0,)$$
leading to the energy-eigenvalue spectrum $\{E_0,E_1,\ldots,E_4\}$
defined as roots of secular equation
 \be
 {{\it E}}^{5}-4\,{{\it E}}^{3}+3\,{\it E}-R=0\,.
 \ee
This may be re-read as an explicit polynomial definition of the
couplings $R=R(E)$ yielding a typical wiggly-curve spectral pattern
as shown in Fig.~\ref{fi05}. The related domain
  ${\cal D}_a$ of couplings $R$
in which the whole spectrum remains real
is an interval   ${\cal D}_a
  =\left (-{R}^{(KEP)}_a,{R}^{(KEP)}_a\right )$
determined by its two KEP
boundaries with
 \ben
 {R}^{(KEP)}_{a}={\frac {12+8\,\sqrt {21}}{125}}\,\sqrt {30-5\,\sqrt
 {21}}
  \approx 1.036340418\,.
 \een
The ground state remains robustly real so that the latter two KEPs
play the role of the points of a merger and of a subsequent
complexification of the first two excitations $E_1$ and $E_2$ and of
the second two excitations $E_2$ and $E_3$ at the left and right end
of the interval, respectively.

%

\begin{figure}[h]                     
\begin{center}                         
\epsfig{file=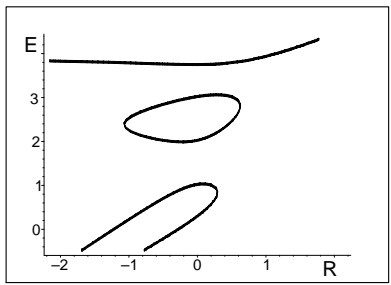,angle=270,width=0.3\textwidth}
\end{center}                         
\vspace{-2mm}\caption{The ${R}-$dependence of energies in potential
(\ref{tripl}) at $\varrho=\varrho_b$.
 \label{fi05b}}
\end{figure}

A different spectral pattern is obtained at
 $$\varrho_b= (1,1,1,1,1) +i\,(0,0,0,0,0,)$$
(see the spectral plot in Fig.~\ref{fi05b}). The complexification
involves there the ground state $E_0$ which merges with the first
excited energy level $E_1$ at the right end of the interval ${\cal
D}_b$. Again, just the two excited energy levels $E_2$ and $E_3$
merge at the left end of physical interval ${\cal D}_b$.

\begin{figure}[h]                     
\begin{center}                         
\epsfig{file=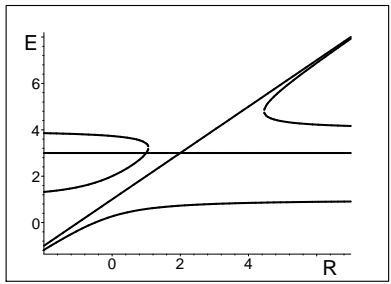,angle=270,width=0.3\textwidth}
\end{center}                         
\vspace{-2mm}\caption{The ${R}-$dependence  of energies for Hamiltonian
(\ref{dvojak}) at $\varrho=\varrho_c$.
 \label{fi06}}
\end{figure}

The domain ${\cal D}$ may cease to be compact and/or simply
connected. Both of these features characterize our third
illustration with
 $$\varrho_c= (1,1,0,0,0) +i\,(0,0,0,0,0,)$$
for which the Hamiltonian
 \be
 H_c(R)=
 \left[ \begin {array}{ccccc}
 2+{R}&-1+{R}&0&0&0
 \\\noalign{\medskip}-1&2&-1&0&0
 \\
 \noalign{\medskip}0&-1&2&-1&0
 \\
 \noalign{\medskip}0&0&-1&2&-1+{R}
 \\
 \noalign{\medskip}0&0&0&-1&2+{R}\end {array}
 \right]
 \label{dvojak}
  \ee
yields, for shifted $x=E-2$, the factorized secular equation
 \be
 \left( {\it x}-1 \right)  \left( {\it x}+1-{R} \right)  \left[
 {{\it x}}^{3}-{{\it x}}^{2}{R}- \left( 3-{R} \right) {\it x}+2\,{R}
 \right]=0\,.
 \label{facto}
 \ee
The spectral pattern becomes anomalous in containing straight lines
(cf. Fig.~\ref{fi06}). The coupling re-expressed as the function of
the energy ${R}(E)$ develops two poles at $x=-1$ and at $x=2$. This
explains why we spot, in the picture, the shared real-constant
asymptotics of $E_2(R)$ and $E_4(R)$ at $ {R} \ll -1$ and of
$E_0(R)$ and $E_2(R)$ at $ {R} \gg 1$,  respectively.

\subsection{$N=6$ example}

\begin{figure}[h]                     
\begin{center}                         
\epsfig{file=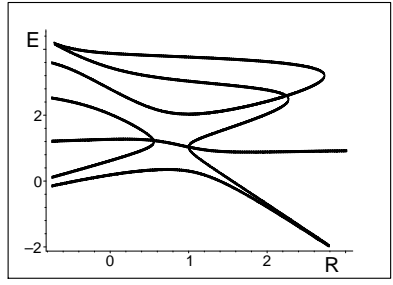,angle=270,width=0.3\textwidth}
\end{center}                         
\vspace{-2mm}\caption{The ${R}-$dependence of the real eigenvalues
of Hamiltonian (\ref{dvojaky}).
 \label{fi068}}
\end{figure}

For the $c-$subscripted $N=5$ Hamiltonian (\ref{dvojak}) with
${R}=1$ it is straightforward to verify that the unavoided
energy-level crossing $E_2(R) \to E_3(R) \to 3$ is accompanied by
the parallelization of the two related eigenvectors. In other words,
the canonical form of the Hamiltonian will contain a two-dimensional
Jordan block at the ${R}=1$ KEP singularity. From the purely
phenomenological point of view such a possibility represents an
independent and important merit of the model which would certainly
deserve a deeper systematic study at a general dimension $N$. Before
we show in which way such a study gets significantly facilitated by
the sparse-matrix nature of our present ${\cal PT}-$symmetric
Hamiltonian~(\ref{HaKa7}), let us still reanalyze another, slightly
more complicated quantum system described by one of the simplest
non-tridiagonal toy-model Hamiltonians
 \be
 H^{(7)}(R)=\left[ \begin {array}{ccccccc}
 2-R&{\it R-1}&{\it R}&0&0&0&0
 \\
 \noalign{\medskip}{\it R-1}&2&-1&0&0&0&0
 \\
 \noalign{\medskip}0&-1&2&
 -1&0&0&0
 \\
 \noalign{\medskip}0&0&-1&2&-1&0&0
 \\
 \noalign{\medskip}0&0&0&-1
 &2&-1&{\it R}
 \\
 \noalign{\medskip}0&0&0&0&-1&2&{\it R-1}
 \\
 \noalign{\medskip}0&0&0&0&0&{\it R-1}&2-R
 \end {array} \right]\,.
 \label{dvojaky}
 \ee
This enables us to demonstrate that and how the use of more levels
(here, $N+1=7$) and/or of more non-vanishing matrix elements in $V$
(here, $M(V)=8$) may make the spectrum increasingly complicated and
richer (cf. Fig.~\ref{fi068}).



\section{Construction of bound states \label{solgiga}}

\subsection{Separable-interaction reinterpretation of $V$}

In the Dirac's bra-ket notation let us denote the elements of the
orthonormalized basis in which we work by  ket-symbols $|e_k\kt$
with property ${\cal P}|e_0\kt = |e_N\kt$. This enables us to
rewrite our ${\cal PT}-$symmetric matrix $V=H-T$ of
Eq.~(\ref{HaKa7}) in the form of a sum
 \be
 V=V_z+V_{ab}+V_{\alpha}+V_{ \beta}\,
 \ee
of four components containing two dyadic terms each,
 \ben
 V_z=
 - z\, |e_0\kt \br e_0| -  z^*\, |e_N\kt \br e_N|\,,
 \ \ \ \
 V_{ab}= a\,
 |e_N\kt \br e_0| + b\, |e_0\kt  \br e_N|\,,
 \een
 \ben
 V_{\alpha }=
 |\vec{\alpha}\kt  \br e_0| + |e_N\kt  \br \vec{\alpha}|\,{\cal
 P}^{(N-1)}\,,
 \ \ \ \
 V_{\beta}=
 |\vec{\beta}\kt  \br e_N| + |e_0\kt  \br \vec{\beta}|\,{\cal
 P}^{(N-1)}\,.
 \een
It is easy to verify that our operator $V$
satisfies the PTS {\em alias} PHR relation
 \be
 V^\dagger{\cal P}^{(N+1)}={\cal P}^{(N+1)}\,V\,.
 \label{ptsymm}
 \ee
What is more important is that the use of projector
 \ben
 \Pi=\sum_{j=1}^{N-1}
 |e_j\kt \br e_j|
 \een
enables us to split
Schr\"{o}dinger
equation into its two boundary layers and a $\Pi-$projected
subsystem,
 \be
 \left(
 \begin{array}{c|ccc|c}
 -z&&\br \beta|{\cal P}^{(N-1)}&&b\\
 \hline
 &0&\ldots&0&\\
 |\alpha \kt&\vdots&\ddots&\vdots&|\beta\kt\\
 &0&\ldots&0&\\
 \hline
 a&&\br \alpha |{\cal P}^{(N-1)}&&-z^*
 \ea
 \right ) \left (
 \ba
 x_-\\
 |X\kt\\
 x_+
 \ea
 \right )=
 \left(
 EI^{(N+1)}-T^{(N+1)}
 \right )
 \left (
 \ba
 x_-\\
 |X\kt\\
 x_+
 \ea
 \right )
 \,.
 \ee
We separated the bulk wave function (i.e., its $N-1$ inner
components $|X\kt= \Pi\,|\psi \kt$) from its values $x_\mp=\br
e_{0,N}|\psi \kt$ at the respective spatial boundaries. In terms of
an $(N-1)-$dimensional and, presumably, invertible matrix
 \be
 \ \ \ \ {\Sigma}^{-1}(E)=\Pi\,\left(
 EI^{(N+1)}-T^{(N+1)}
 \right )\Pi
 \label{weemp}
 \ee
we may separate the bulk part of our Schr\"{o}dinger equation, i.e.,
its $N-1$ inner lines
 \ben
 |\alpha \kt\,x_-+|\beta\kt\,x_+
 = {\Sigma}^{-1}(E)\,
  |X\kt+
 |e_1\kt \,x_-+|e_{N-1}\kt \,x_+
 \,.
 \een
They may be formally multiplied by the matrix resolvent
${\Sigma}(E)$ from the left yielding the $\Pi-$projection $|X\kt$ of
our wave function in the manifestly $E-$dependent explicit form
 \be
 |X\kt =
 {\Sigma}(E)\, \left (
  |\alpha \kt-|e_1\kt
  \right )\,x_-
  + {\Sigma}(E)\,
  \left (|\beta\kt-|e_{N-1}\kt
  \right )\,x_+=|A_-(E)\kt x_-+|B_+(E)\kt x_+
  \,.
  \label{vectro}
 \ee
Due to our discrete-Laplacian choice of matrix $T$ the matrix
inversion ${\Sigma}^{-1}(E) \to {\Sigma}(E)$ is non-numerical (see
the next subsection). Hence, vector (\ref{vectro}) is known in
closed form and it may be inserted in the remaining two lines of
Schr\"{o}dinger equation. We end up with the two homogeneous linear
equations
 \be
 (2-E-z)\,x_-+
 \br \beta |{\cal P}^{(N-1)}|X\kt -\br e_1|X\kt
 +b\,x_+=0\,,
 \ee
 \be
 a\,x_-+
 \br \alpha |{\cal P}^{(N-1)}|X\kt -\br e_{N-1}|X\kt
 +(2-E-z^*)\,x_+=0\,
 \ee
determining the last two missing components $x_\pm$ of the wave
function. Naturally, the set of bound state energies $E$ is finally
obtained as zeros of the related secular two-by-two determinant.

\subsection{Non-numerical diagonalization and construction of
$\Sigma(E)$}

Undoubtedly, the exact solvability of our present model is due to
the non-numerical invertibility of the $(M+1)$ by $(M+1)$ matrices
(\ref{weemp}) at any dimension $M+1$. More details may be found
elsewhere \cite{square}. For our present purposes it is sufficient
to remind the readers that the Chebyshev polynomials $U_n(y)$ of the
second kind satisfy the relation \cite{cheby}
 \be
 \left[ \begin {array}{ccccc}
 2{y}&-1&0&\ldots&0
 \\
 -1&2{y}&\ddots&\ddots& \vdots \\
 0&-1&\ddots&-1&0 \\
 \vdots&\ddots&\ddots&2{y}&-1\\
 0&\ldots&0&-1&2{y}\\
  \end {array} \right]\,
   \left ( \ba
    U_0(y)\\
    U_1(y)\\
    \vdots\\
    U_M(y)
    \ea
    \right ) = 0
    \label{colu}
    \ee
provided only that we choose the value of $y$ as one of the $M+1$
roots of the polynomial equation $U_{M+1}(y)=0$, i.e.,
 \be
 y=y_j^{(M+1)}=2 \cos \frac{(j+1)\pi}{M+2}\,,
 \ \ \ \ \ j = 0, 1, \ldots, M
 \,.
 \ee
As long as all of these roots are known in closed form, we may
arrange the $(M+1)-$plet of the column-vector solutions of
Eq.~(\ref{colu}) into a unitary matrix ${\cal U}$ and diagonalize
the discrete-Laplacean matrix $T^{(M+1)}$ since
 \be
 \left[ \begin {array}{ccccc}
 2{}&-1&0&\ldots&0
 \\
 -1&2{}&\ddots&\ddots& \vdots \\
 0&-1&\ddots&-1&0 \\
 \vdots&\ddots&\ddots&2{}&-1\\
 0&\ldots&0&-1&2{}\\
  \end {array} \right]
 ={\cal U}\,
 \left[ \begin {array}{ccccc}
 {\delta}_0^{(M+1)}{}&{0}&0&\ldots&0
 \\
 {0}&{{\delta}_1^{(M+1)}}{}&\ddots&\ddots& \vdots \\
 0&{0}&\ddots&{0}&0 \\
 \vdots&\ddots&\ddots&{{\delta}_{M-1}^{(M+1)}}{}&{0}\\
 0&\ldots&0&{0}&{{\delta}_M^{(M+1)}}{}\\
  \end {array} \right]\, {\cal U}^\dagger\,
  \label{7}
 \ee
with $\delta^{(M+1)}_j=2\,\left (1-y^{(M+1)}_j\right )$. Once we now
choose $M=N-2$, it becomes easy to return to Eq.~(\ref{weemp}) and
to perform, non-numerically, the inversion ${\Sigma}^{-1}(E) \to
{\Sigma}(E)$.

\section{Construction of physical metrics\label{secty}}


Several detailed studies \cite{Zelezny} confirmed, by non-numerical
constructive means, that in an {\it ad hoc} physical Hilbert space
${\cal H}^{(S)}$ the evolution of the differential-equation model
(\ref{SE}) +  (\ref{spojhambcdavid}) can be made unitary. In
Ref.~\cite{I}, analogous results were obtained for the
difference-equation model (\ref{DSE}) + (\ref{RBC}). In both of the
latter contexts the explicit construction of the physical Hilbert
space ${\cal H}^{(S)}$, i.e., of its inner-product metric operator
$\Theta$ represented not only the quantum-theoretical necessity but
also by far the most difficult mathematical challenge.

\subsection{Expansions of $\Theta$ in terms of eigenvectors of
$H^\dagger$}

In technical terms one has
to find such an operator $\Theta=\Theta^\dagger>0$ which
is connected with the given Hamiltonian $H\neq H^\dagger$ via
the $S-$space Hermiticity
condition
 \be
 H^\dagger\Theta=\Theta\,H\,.
 \label{Dieudonnes}
 \ee
In an abstract context of pure mathematics
such an $S-$space Hermiticity relation
was
studied by Dieudonn\'{e} \cite{Dieudonne}.
More or less simultaneously
the same generalization of Hermiticity
proved useful for
physicists
who found
a number
of its useful practical applications in many-body quantum
systems \cite{Geyer,Dyson}.
In what follows we shall show that
the construction of a suitable Hermitizing
metric $\Theta$
remains feasible
for
our present Hamiltonians (\ref{HaKa7}).

For finite-dimensional Hamiltonians there exist two alternative and
efficient methods of the reconstruction of the necessary Hermitizing
metric $\Theta=\Theta(H)$. The more universal spectral-expansion
approach as discussed in Ref.~\cite{SIGMAdva} is based on formula
 \be
 \Theta=\sum_0^N\,|n\kkt\kappa^2_n\bbr n|\,,
 \ \ \ \ \kappa_n \in \mathbb{R}\setminus \{0\}\,.
 \label{sere}
 \ee
The special kets $|n\kkt$ are defined here as a complete
set of solutions of
the conjugate-Hamiltonian Schr\"{o}dinger equation
 \be
 H^\dagger\,|n\kkt = E_n\,|n\kkt\,,
  \ \ \ \ \
  n = 0, 1, \ldots, N\,.
  \label{conse}
  \ee
In general, one merely has to employ  the method of solution as
described in section \ref{solgiga} above. Alternative versions of
the method may be developed for some simpler special cases. One of
them will be described in what follows.

\subsection{Closed-form metrics for $|\vec{\beta}\kt = \vec{0}$}

Let us restrict our attention to the subset of models (\ref{HaKa7})
with trivial $|\vec{\beta}\kt = \vec{0}$, i.e., with the conjugate
Schr\"{o}dinger operator
 \be
 \left (H^{(N+1)}_0-E\,I^{(N+1)}
 \right )^\dagger
 =\left[ \begin {array}{cccccc|c}
  2{y}-z^*&{\alpha}_{1}^*-1&\alpha^*_{2}&
  \ldots&\alpha_{N-2}^*&\alpha_{N-1}^*&a
 \\
 \hline
 -1&2{y}&-1&0&\ldots&0&\alpha_{N-1}
 \\
 0&-1&2{y}&\ddots&\ddots& \vdots&\alpha_{N-2} \\
 0&0&-1&\ddots&-1&0&\vdots \\
 \vdots&\vdots&\ddots&\ddots&2{y}&-1&\alpha_{2}\\
 0&0&\ldots&0&-1&2{y}&\alpha_{1}-1\\
 0&0&0&\ldots&0&
 -1&2{y}-z
  \end {array} \right]
  \,
 \label{HuKu7}
 \ee
where we abbreviated $2-E=2y$ and partitioned
 \be
 \left (H^{(N+1)}_0-E\,I^{(N+1)}
 \right )^\dagger
 =
  \left[ \begin {array}{c|c}
  \br \vec{u}(y)|&a\\
  \hline
  -S^{-1}(y)&{\cal P}^{(N)} |\vec{u}(y)\kt
  \end {array} \right]\,.
 \ee
We insert this matrix in  Eq.~(\ref{conse}) and get a re-partitioned
conjugate Schr\"{o}diner equation
 \be
  \left[ \begin {array}{cc}
  \br \vec{u}(y_n)|&a\\
  -S(y_n)&{\cal P}^{(N)} |\vec{u}(y_n)\kt
  \end {array} \right]\,|n\kkt=0\,,
  \ \ \ \ \
  |n\kkt=\left(
  \ba
  |\chi_n\kt\\
  x_n
  \ea
  \right)
  \,.
  \label{Isabel}
 \ee
As long as matrices $H$ and $H^\dagger$ are isospectral, the
bound-state energy $E_n$ (together with its two reparametrizations
$2y_n=2-2E_n=t_n+1/t_n$) is already known, at any $n=0,1,\ldots,N$,
from section \ref{solgiga}. Thus, the first line of
Eq.~(\ref{Isabel}) may be omitted as redundant. In terms of an
arbitrary normalization constant $x_n$ the rest of this equation
determines the ultimate $N-$dimensional vector
 \be
  |\chi_n\kt=S^{-1}(y_n)\,{\cal P}^{(N)} |\vec{u}(y_n)\kt\,x_n\,.
  \label{forje}
  \ee
It is important that the explicit form of the inverse of our
triangular $N$ by $N$ auxiliary matrix is obtainable in closed form,
 \be
 S^{-1}(y_n)=U(t_n)U(1/t_n)\,,
 \ \ \ \ U(\tau)=\left[ \begin {array}{ccccc}
 1&\tau&\tau^2&\ldots&\tau^{N-1}\\
 0&1&\tau&\ldots&\tau^{N-2}\\
 0&0&1&\ddots&\vdots\\
 \vdots&\ddots&\ddots&\ddots&\tau\\
 0&0&\ldots&0&1
  \end {array} \right]\,.
  \label{fordor}
 \ee
Thus, the insertion of Eqs.~(\ref{forje}) and (\ref{fordor}) in
Eq.~(\ref{sere}) completes the construction of the metric. We see
that due to the simplification  $|\vec{\beta}\kt = \vec{0}$ we did
not have to pre-diagonalize and invert the tridiagonal
kinetic-energy matrix $T$ at all.

\section{Sparse-matrix metrics\label{sepet}}

The key advantage of the strategy of the preceding section may be
seen in an immanent guarantee of the positivity of the metric. A
weak point of the method is that even if the Hamiltonian itself
admits an exceptionally simple (i.e., diagonal or other
sparse-matrix) form of the metric $\Theta=\Theta(H)\neq I$, it is
difficult to identify such a most welcome solution of
Dieudonn\'{e}'s Eq.~(\ref{Dieudonnes}) as a special case of the
general spectral expansion (\ref{sere}). A different construction
strategy is needed in such a situation.

\subsection{Recurrent constructions}

In the spectral-expansion formula (\ref{sere}) the
physics-determining choice (i.e., the
$S-$superscripted-Hilbert-space-determining choice) of the
$(N+1)-$plet of positive parameters $\kappa_n^2$ is {\em fully} at
the user's disposal (cf. Ref.~\cite{Geyer} for an exhaustive
analysis and explanation and physical interpretation of such an
apparent ambiguity). Naturally, in an alternative approach one may
choose free parameters directly as some matrix elements of $\Theta$.
Then, one has to determine the remaining elements of the Hermitizing
metric $\Theta$ via a direct, brute-force solution of the linear
algebraic system of $(N+1)^2$ equations (\ref{Dieudonnes}) (which
are not always mutually independent).

An efficient implementation of the latter recipe is much less
universal and its feasibility depends strongly on the structure and
properties of the input Hamiltonian $H \neq H^\dagger$. For
illustration purposes let us consider, therefore, just the
one-parametric five-dimensional real-matrix Hamiltonian $H_c(R)$ of
Eq.~(\ref{dvojak}). In accord with Eq.~(\ref{facto}) and
Fig.~\ref{fi06} we know that the energies are all real, say, at any
$R<1$. It is also easy to prove that the simplest possible
Hermitizing metric  $\Theta^{(c)}>0$ may be chosen as a diagonal
matrix with elements $\Theta_{00}^{(c)}=1$,
$\Theta_{11}^{(c)}=\Theta_{22}^{(c)}=\Theta_{33}^{(c)}=1-{R}$ and
$\Theta_{44}^{(c)}=(1-{R})^2$. Nevertheless, it would be fairly
difficult to derive this result from the mere inspection of spectral
expansion (\ref{sere}). An alternative, recurrent recipe may be
recommended.

\begin{lem}

\label{solemnis}

The complete family of the real and symmetric matrices $\Theta$
which Hermitize
Hamiltonian
$H_c(R)$
of Eq.~(\ref{dvojak})
may be constructed
by solving Eq.~(\ref{Dieudonnes}) in recurrent manner.

\end{lem}

\bp

Let us treat matrix elements $u{}=\Theta_{01}$, $z{}=\Theta_{02}$,
$q{}=\Theta_{03}$, $p{}=\Theta_{04}$ and $t{}=\Theta_{22}$ as a
quintuplet of independently variable lower-case parameters in the
metric. The subsequent inspection of relations (\ref{Dieudonnes}),
viz.,
 \ben
 {\cal M}_{ij}=
 \sum_{k=0}^4\,\left (H^\dagger\right )_{ik}
 \Theta_{kj}-\sum_{n=0}^4\,\Theta_{in}
 H_{nj}=0
 \een
reveals that their $_{ij}={}_{04}$ item defines
$\Theta_{14}=q{}-R\,q{}$, their $_{ij}={}_{03}$ item defines
$\Theta_{13}=R\,q{}+z{}+p{}$ and their $_{ij}={}_{02}$ item defines
$\Theta_{12}=R\,z{}+u{}+q{}$. Next, after insertions of these
elements their $_{ij}={}_{14}$ item defines $\Theta_{24}=z{}-R\,z{}$
and their $_{ij}={}_{13}$ item defines $\Theta_{23}=q{}+R\,z{}+u{}$.
Next, after insertions of these elements their $_{ij}={}_{24}$ item
defines $\Theta_{34}=u{}-R\,u{}$. Next, after the insertion of this
element their $_{ij}={}_{12}$  item defines diagonal element
$\Theta_{11}=-R\,q{}-p{}+t{}-R\,z{}$ and their $_{ij}={}_{23}$  item
defines diagonal element $\Theta_{33}=-R\,q{}-p{}+t{}-R\,z{}$.
Finally, after insertions of these elements their $_{ij}={}_{01}$
item defines $\Theta_{00}=(R\,z{}-t{}+z{}+R\,u{}+R\,q{}+p{})/(-1+R)$
and their $_{ij}={}_{34}$ item defines $\Theta_{44}=-z{}
 +R\,p{}-R\,q{}+R^2\,q{}-p{}-R\,u{}+t{}-t{}\,R+R^2\,z{}+R^2\,u{}$.

\ep

In Fig.~\ref{fi06} we may notice (and, if asked for, we may readily
prove) that for Hamiltonians $H_c(R)$ the energy spectrum is real in
the interval of $R \in (-\infty,R_{max})$ with boundary value
$R_{max}\approx 1.065260704$ which is larger than one. At the same
time, a return to the proof of the preceding Lemma implies that
every candidate $\Theta=\Theta(t,u,z,q,p)$ for the metric (say, with
$t=1$) is a diagonal matrix if and only if one sets $u=z=q=p=0$. The
resulting matrix $\Theta(1,0,0,0,0)$ is positive definite (i.e., it
may serve as a physical metric) if and only if $R  < 1$. This is an
incompatibility which has the following important physical
interpretation.

\begin{lem}

If we assign the same diagonal metric $\Theta(1,0,0,0,0)$ to
Hamiltonian $H_c(R)$ at all $R<1$, the passage through $R=1$ will
necessarily lead to a phase transition, {\em not accompanied} by the
loss of observability of the bound-state energy spectrum.

\end{lem}

\bp

Firstly, it is obvious that the value of $R=1$ is a KEP singularity
of matrix $H_c(R)$. At this point, secondly, the diagonal physical
metric ceases to be positive definite. We may conclude that for
$R\in (1,R_{max})$, {\em any} acceptable metric
$\Theta=\Theta(t,u,z,q,p)$ becomes {\em necessarily} non-diagonal.
To the left and to the right from the KEP singularity $R=1$, the
metrics $\Theta$ (i.e., the physical Hilbert spaces ${\cal
H}^{(S)}$) are necessarily different so that also the quantum system
in question {\em must} be {\em necessarily} characterized by the
{\em different} complete sets of the operators of observables.

\ep

At the right end of the smaller physical interval of $R\in
(1,R_{max})$ the mathematical as well as physical nature of the KEP
singularity is less unusual since at this value the two highest
energies cease to be real. They merge and, subsequently, complexify.
The merger value of energy $E=x+2$ may be determined as a root of
quartic polynomial ${{\it x}}^{4}-2\,{{\it x}}^{3}-3\,{{\it
x}}^{2}+6=0$. Numerically one reveals that $E_3 \to 3.233152848
\leftarrow E_4$ in the limit of $R\to R_{max}$.

\subsection{Diagonal metric candidates}

The recurrent construction method as sampled in preceding subsection
leads to the metric-operator candidates $\Theta$ in which certain
merits (e.g., sparsity) are combined with shortcomings (e.g., a
possible premature loss of the necessary positivity). Naturally, at
any dimension $N+1$ the merits get maximal and, simultaneously, the
shortcomings are minimal for diagonal matrices
$\Theta=\Theta^{(diag)}$ so their subset deserves an enhanced
attention.

At any matrix dimension, the same partitioning may be applied to
Hamiltonians $H^{(N+1)}= T^{(N+1)} +
V^{(N+1)}(z,a,b,\vec{\alpha},\vec{\beta})$ of Eq.~(\ref{HaKa7}) as
well as to the related diagonal metric candidates. Once we separate
the outer elements we may abbreviate $\Theta^{(diag)}_{00}=m_0$ and
$\Theta^{(diag)}_{NN}=m_N$ and use the subscripts $j=1,2,\ldots,N-1$
numbering just the inner part of the diagonal
$\Theta^{(diag)}_{jj}=d_j$ forming a smaller, tilded diagonal matrix
$\tilde{d}$.

We shall assume that the Hamiltonians as well as the metric
candidates are real matrices. In the light of the $N=4$ results of
the preceding subsection we feel encouraged to classify, at all $N$,
the scenario in which the two matrices remain compatible with the
hidden Hermiticity constraint (\ref{Dieudonnes}).

First of all let us restrict our attention to the inner, $(N-1)$ by
$(N-1)$ matrix partition of Eq.~(\ref{Dieudonnes}). The potential
cannot contribute so we must set $d_1=d_2$, \ldots,
$d_{N-2}=d_{N-1}$. Without any loss of generality we have
$d_{j}=d>0$ at all $j$, i.e.,  $\tilde{d}=d\,I^{(N-1)}$.

After we turn attention to the four outer partitions of
Eq.~(\ref{Dieudonnes}), we may study just its first and last columns
of the real and symmetric matrix set ${\cal M} = H^\dagger
\Theta-\Theta\,H=0$ with a trivial main diagonal. The first column
offers a set of $N$ relations
 \be
 ({\beta}_{N-1}-1)\,m_0=d\,( {\alpha}_1-1)\,,
 \ \ \ \
 {\beta}_{N-j}\,m_0=d\, {\alpha}_{j}\,,
 \ \ \ j = 2, 3, \ldots, N-1\,,
 \ \ \ \
 b\,m_0=m_N\,a\,.
 \label{under}
 \ee
They have to be complemented by the  $N-$plet of the
last-column relations
 \be
 a\,m_N =m_0\,b\,,
 \ \ \ \
 {\alpha}_{N-j}\,m_N=d\, {\beta}_{j}\,,
 \ \ \ j = 1, 2, \ldots, N-2\,,
 \ \ \ \
 ({\alpha}_{1}-1)\,m_M=d\,( {\beta}_{N-1}-1)\,
  \ee
in which, due to the symmetry of ${\cal M}$, the first item is
already redundant of course.

As long as we must have positive $d > 0$ and $m_N>0$ for any
diagonal metric, the former set of relations may be now read as the
definition of the $N-$plet of ``acceptable'' parameters $\alpha_j$
and $a$ in terms of the respective ``optional'' parameters
$\beta_{N-j}$ and $b$. In such a case we may insert these quantities
in the latter  set of relations and discover that up to the
uninteresting trivial case (with a completely vanishing
nonlocal-boundary-condition part of the Hamiltonian) the whole
resulting  set of relations degenerates to the single sufficient
condition $m_0\,m_N=d^2$. This completes the proof of the following
result.

\begin{prop}

For the $(N+1)$ by $(N+1)$  real-matrix Hamiltonian (\ref{HaKa7})
there exists a positive-definite diagonal metric $\Theta^{(diag)}$
if and only if the $N-$plets of the dynamical parameters
$\vec{\alpha}$ (and $a$) and $\vec{\beta}$ (and $b$) satisfy the
$N-$plet of linear relations  (\ref{under}) where $d>0$ is a fixed
inessential overall multiplier and where $m_0\,m_N=d^2$.

\end{prop}

\begin{pozn}

While the diagonal metric may vary with just a single ``essential''
free parameter, the variability of the energies remains controlled
by an $N-$plet of free couplings.

\end{pozn}

\section{Parametric domains ${\cal J}$ of positivity of the metric
\label{sepetbe}}

\subsection{$N=4$ example}

A decisive merit of the above-sampled and comparatively complicated
recurrent construction of the complete set of the metric candidates
should be seen in the facilitated construction of their
sparse-matrix subsets and even, in some cases, of their
$(2K+1)-$diagonal band-matrix versions. This is well illustrated by
the above-derived explicit formula
 \be
  \Theta(t,u,z,q,p)=\left[ \begin {array}{ccccc}
  m(00)&{\it u{}}&{\it z{}}&{\it q{}}&{\it p{}}
 \\\noalign{\medskip}{\it u{}}&m(11)&R{
 \it z{}}+{\it u{}}+{\it q{}}
 &R{\it q{}}+{\it z{}}+{\it p{}}&{\it q{}}-R{\it
 q{}}\\\noalign{\medskip}{\it z{}}
 &R{\it z{}}+{\it u{}}+{\it q{}}&{\it t{}}&R
 {\it z{}}+{\it u{}}+{\it q{}}&{\it z{}}-R{\it z{}}
 \\\noalign{\medskip}{\it
 q{}}&R{\it q{}}+{\it z{}}+{\it p{}}&R{\it z{}}+{\it u{}}
 +{\it q{}}&m(33)&{\it u{}}-R{\it u{}}
 \\\noalign{\medskip}{
 \it p{}}&{\it q{}}-R{\it q{}}
 &{\it z{}}-R{\it z{}}&{\it u{}}-R{\it u{}}&
 m(44)
 \end {array} \right]
 \label{metrac}
 \ee
for $H=H_c(R)$ of Eq.~(\ref{dvojak}) where we abbreviated
 \be
 m(00)={\frac {R{\it z{}}-{\it t{}}+{\it z{}}+R{\it u{}}
 +R{\it q{}}+{\it p{}}}{-1+R}}\,,
 \ \ \ \
 m(11)=m(33)=-R{\it q{}}-{\it p{}}+{\it t{}}-R{\it z{}}
 \label{regu}
 \ee
and
 $$
 m(44)=-{\it z{}}+R{\it p{}}-R{\it q{}}+{R}^{2}{\it q{}}
 -{\it p{}}-R{\it u{}}+{\it
 t{}}-{\it t{}}\,R+{R}^{2}{\it z{}}+{R}^{2}{\it u{}}\,.
 $$
Almost all of the matrix elements are just linear functions of the
parameters $t,u,z,q,p$ and $R$. For this reason, the sufficiently
small deviations of the parameters from their diagonal-metric limit
$u=z=q=p=0$ can merely restrict (but not entirely destroy) the
robust positivity (i.e., acceptability) of the diagonal-metric
candidate.

For illustrative purposes let us pay attention just to the first
nontrivial, tridiagonal-matrix special case of Eq.~(\ref{metrac})
and let us re-analyze the one-parametric family of the candidates
$\Theta=\Theta(1-R,u,0,0,0)$ for the metric. Our first observation
is that even at the very small fixed $u$ we cannot exclude
irregularities in the $R \to 1$ limit. The reason is that at any
fixed value of $u>0$ the matrix element $m(00)$ as given by
Eq.~(\ref{regu}) remains singular in the limit, $\lim_{R \to
1^-}m(00)=-\infty$. Near the singular value of $R$ the matrix
element $m(00)$ is dominant and becomes approximatively equal to one
of the eigenvalues. Simultaneously, as long as this matrix element
becomes large and negative for positive $u$ in this limit, the
candidate matrix becomes indeterminate and, hence, inacceptable near
the KEP boundary of $R \approx 1$.

\begin{figure}[h]                     
\begin{center}                         
\epsfig{file=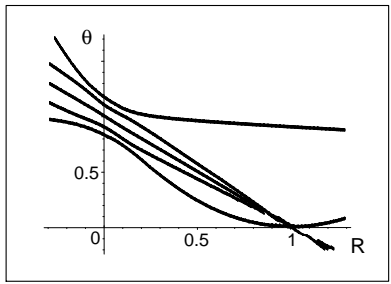,angle=270,width=0.3\textwidth}
\end{center}                         
\vspace{-2mm}\caption{The $R-$dependence of the
 eigenvalues of the tridiagonal $N=5$ matrix $\Theta(t,u,0,0,0)$
 of Eq.~(\ref{metrac}) at
 normalization $t=1-R$ and small $u=(1-R)/10$.
 \label{fi08}}
\end{figure}

%
%

 \begin{figure}[h]                     
 \begin{center}                         
 \epsfig{file=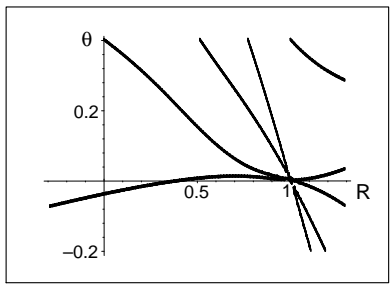,angle=270,width=0.3\textwidth}
  \end{center}                         
 \vspace{-2mm}\caption{The same ${R}-$dependence as in Fig.~\ref{fi08}
  but at almost critical  $u=6(1-R)/10$.
 \label{fi09}}
 \end{figure}


Fortunately, after one regularizes the $R=1$ singularity in $m(00)$,
the rest of the metric candidate already appears to be a smooth
function of its parameters. At their small values one may expect
just small deviations of eigenvalues from their unperturbed limit
with $\theta_0=1$, with the degenerate triplet of the linearly
$R-$dependent  eigenvalues $\theta_{1,2,3}=1-R$ and with the
quadratically $R-$dependent eigenvalue $\theta_4=(1-R)^2$.

Once we consider just the first nontrivial and properly regularized
tridiagonal matrix $\Theta=\Theta(1-R,(1-R)\,w,0,0,0)$, our
expectation of the smoothness of the variation of the eigenvalues
are confirmed. This is documented by Fig.~\ref{fi08}  using a small
$w=1/10$. We see there that whenever $R< 1$ the spectrum of the
metric remains positive. As expected, its $R-$dependence keeps the
clear trace of the separation of one eigenvalue of an almost
constant type from an almost linear  triplet and from an almost
quadratic shape, at the not too small values of $R$ at least. Only
far to the left from $R=1$ (i.e., near $R=0$) this pattern changes
since avoided crossings enter the scene.


 \begin{figure}[h]                     
 \begin{center}                         
 \epsfig{file=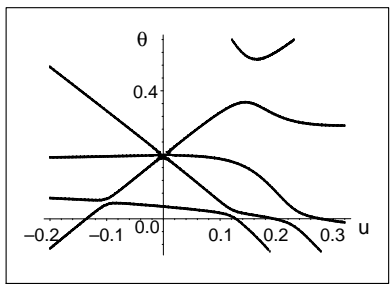,angle=270,width=0.3\textwidth}
  \end{center}                         
 \vspace{-2mm}\caption{The ${u}-$dependence of the
 eigenvalues of $\Theta(1-R,u,0,0,0)$
 at
  $R=8/10$.
 \label{fi09b}}
 \end{figure}

 %

With the growth of $w$ towards a certain not too small
$w=w_{critical}$ we expect that at least one of the eigenvalues of
the tridiagonal metric candidate $\Theta(1-R,(1-R)\,w,0,0,0)$ will
vanish so that beyond the boundary point $w_{critical}$ the
candidate matrix will become indeterminate (i.e., for our present
purposes, useless). Recalling and comparing Figs.~\ref{fi08}
and~\ref{fi09} we may visualize the empirical observation that the
point $R_{min}$ of the loss of the positivity of the metric
candidate moves up with the growth of $w$.

Using the brute-force numerics we managed to arrive at a rough
estimate of the maximal possible $w=w_{critical} \in (7/10, 8/10)$.
Nevertheless, after a return to Fig.~\ref{fi09} one sees that in a
comparatively large vicinity of $w_{critical}$ the minimal positive
eigenvalue $\theta_0$ is in fact rather small in the whole
admissibility interval of $R \in (R_{min},1)$. Thus, the numerical
invertibility of the metric would certainly deteriorate long before
the value of $w$ reaches its critical value. Thus, the model will
only prove numerically well conditioned well below the maximal
$w_{critical}$.

The regularization $u \to w$ is not needed at a fixed dynamical
coupling $R<1$. We choose $R=4/5$ and displayed, in
Fig.~\ref{fi09b}, the graphical proof of the positivity of all of
the eigenvalues $\{\theta_j\}$ of the tridiagonal metric candidate
$\Theta(1/5,u,0,0,0)$ in an interval of the length which is not too
small (i.e., in fact, it is bigger than $1/5$).

\subsection{Arbitrary dimensions $N+1$ \label{arbi}}

In Ref.~\cite{recur} it has been shown that the recurrent
constructions of the metric candidates remain particularly efficient
for the sparse-matrix Hamiltonians which are at most tridiagonal.
Hence, in a search for a more challenging/rewarding illustrative
example let us now return, once more, to the asymmetric and
non-tridiagonal real-matrix Hamiltonian $H^{(7)}(R)$ of
Eq.~(\ref{dvojaky}). For this model we know that the spectrum is
real in Hermitian limit $R\to 0$ \cite{square} and that it remains
all real at small $R \lessapprox 0.555$ at least (cf.
Fig.~\ref{fi068} above). This encourages us to formulate and prove
the following generalization of lemma \ref{solemnis}.

\begin{prop}

\label{solemnisbe}

The complete family of the real and symmetric matrices
$\Theta^{(N+1)}$ which pseudo-Hermitize the minimally
nonlocal-boundary $M(V)=8$ Hamiltonians $H^{(N+1)}$ as sampled by
matrix $H^{(7)}(R)$ of Eq.~(\ref{dvojaky}) may be constructed by
solving pseudo-Hermiticity condition~(\ref{Dieudonnes}) in recurrent
manner.

\end{prop}

\bp

Proof proceeds in a complete parallel to the proof of lemma
\ref{solemnis}. It starts by the interpretation of the first row of
matrix elements $x_1=\Theta_{00}$, $x_2=\Theta_{01}$, \ldots,
$x_{N+1}=\Theta_{0N}$ (one has, in our new illustrative example,
$N=6$) as independent parameters. After the explicit display of
relations (\ref{Dieudonnes}) in matrix form
 \ben
 {\cal M}_{ij}= \sum_{k=0}^N\,\left (H^\dagger\right )_{ik}
 \Theta_{kj}-\sum_{n=0}^N\,\Theta_{in}
 H_{nj}=0
 \een
one has to proceed as above. Thus, the selection of the set of
independent equations starts from ${\cal M}_{0N}=0$ and moves to the
left up to the diagonal element which is found trivial and omitted.
In the next round one moves down and repeats the selection. In this
manner one obtains the sequence of $(N+1)N/2$ linear equations
${\cal M}_{ij}=0$ (with $i<j$) for single unknown quantity yielding
easily, with insertions performed after completion of each row, the
sequence of the required explicit formulae for the respective matrix
elements $\Theta_{i+1,j}(x_1,x_2,\ldots,x_{N+1})$.

\ep

As a result of the application of the above algorithm one has the
complete, multiparametric set of matrix candidates
$\Theta^{(N+1)}(x_1,x_2,\ldots,x_{N+1})$ for the metric. All is now
prepared for the final step of the construction of the metric. It is
sufficient just to specify a non-empty domain ${\cal J}$ of
parameters $x_1,x_2,\ldots,x_{N+1}$ (i.e., of vectors $\vec{x}$) in
which the acceptability (i.e., positivity) of the candidate matrix
$\Theta$ is guaranteed.

At all $N$ and at all of our present nonlocal boundary conditions
the latter task is facilitated by the fact that at $z=0$ and in the
Hermitian limit of $\vec{\alpha}\to \vec{0}$ and $\vec{\beta}\to
\vec{0}$ the Hamiltonians degenerate to the discrete Laplacean
$T^{(N+1)}$ with real and non-degenerate spectrum. This means that
we may choose a trivial metric $\Theta=I$ in this limit. Next, in a
sufficiently small vicinity of the values of $z=0$, of
$\vec{\alpha}=\vec{0}$ and of $\vec{\beta}= \vec{0}$ we are sure
that the energy spectrum of $H$ remains, for continuity reasons,
real and discrete (indeed, the complexification may only proceed via
a smoothly proceeding merger of at least one energy doublet
\cite{ali}). By construction and for similar reasons, also the
Hermitizing metrics (with elements being linear functions of
parameters $\vec{x}$) will deviate from the unit matrix while still
remaining positive definite in a suitable non-empty domain  ${\cal
J}$. We just completed the proof of the following result.

\begin{thm}

\label{solemnisbeCE}

At the sufficiently small dynamical parameters $z$, $\vec{\alpha}$
and $\vec{\beta}$ the metric candidates $\Theta^{(N+1)}(\vec{x})$ of
proposition \ref{solemnisbe} constructed as perturbations of
$\Theta^{(N+1)}(\vec{0})=I$ will be positive definite (and will
Hermitize the Hamiltonians) in a non-empty domain ${\cal J}$ of
sufficiently small parameters  $\vec{x}$.

\end{thm}

 \begin{figure}[h]                     
 \begin{center}                         
 \epsfig{file=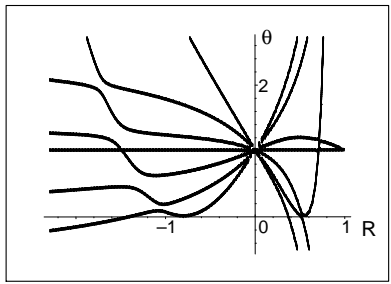,angle=270,width=0.3\textwidth}
  \end{center}                         
 \vspace{-2mm}\caption{The ${R}-$dependence of the
 eigenvalues of metric candidate (\ref{vejir}).
 \label{fi010}}
 \end{figure}

%

In order to illustrate the latter result let us pick up again
Hamiltonian $H^{(7)}(R)$ of Eq.~(\ref{dvojaky}) and using the
algorithm of Proposition \ref{solemnisbe} let us construct the
following illustrative sparse-matrix candidate for the metric,
 \be
 \Theta^{(7)}(1,0,0,0,0,0,0)=
 \left[ \begin {array}{ccccccc}
 1&0&0&0&0&0&0
 \\\noalign{\medskip}0&1&{
 \frac {R}{R-1}}&0&0&0&0
 \\\noalign{\medskip}0&{\frac {R}{R-1}}&1&{\frac {R}{R-1}}&0&0&0
 \\\noalign{\medskip}0&0&{\frac {R}{R
 -1}}&1&{\frac {R}{R-1}}&0&0
 \\\noalign{\medskip}0&0&0
 &{\frac {R}{R-1}}&1&{\frac {R}{R-1}}&-{\frac {{R}^{2
 }}{R-1}}
 \\\noalign{\medskip}0&0&0&0&{\frac {R}{R-1}}&1+R^2
 &-{\frac {{R}^{3}+2\,{R}^{2}-2\,R}{R-1}}
 \\\noalign{\medskip}0&0&0&0
 &-{\frac {{R}^{2}}{R-1}}&-{\frac {{R}^{3}+2
 \,{R}^{2}-2\,R}{R-1}}&{\frac {2\,{R}^{4}-2\,R+1}{{R}^{2}-2\,R+1}}
 \end {array} \right]
 \label{vejir}
 \ee
The variations of its eigenvalues with $R$ are displayed in
Fig.~\ref{fi010}. One may conclude that besides the required
Hermitian-limit property $\lim_{R\to
0}\Theta^{(7)}(1,0,0,0,0,0,0)=I^{(7)}$, the matrix remains also
safely positive definite in a not too small vicinity of the
Hermitian-Hamiltonian limiting point $R=0$.

Another, different lesson taught by the latter example is that at
any dimension $N$, the breakdown of the tridiagonality of the
Hamiltonian may make the structure of the metrics less predictable.
In a way guided by our $N=6$ illustrative $\Theta_{jk}$ one may
expect, for example, that even in the maximally sparse cases the
metrics may not possess the few-diagonal matrix structure anymore.

\section{Pseudometrics \label{pst}}

The pseudo-Hermiticity constraint
 \be
  H^\dagger{\cal P}={\cal P}H
  \label{Dieu}
  \ee
is formally equivalent to the hidden Hermiticity constraint
(\ref{Dieudonnes}), with two differences. Firstly, the matrix of
metric $\Theta$ in Eq.~(\ref{Dieudonnes}) is {\em always} (perhaps,
just tacitly) assumed positive definite and, secondly, it is allowed
to be Hamiltonian-dependent. In contrast, the (invertible) operator
of pseudometric ${\cal P}$ in Eq.~(\ref{Dieu}) is usually required
Hamiltonian-independent and indefinite (plus, sometimes, such that
${\cal P}^2=I$).

Without taking these differences into account, it is obvious that
among the sparse-matrix metric candidates one should find {\em all}
of the indefinite pseudometrics. In particular, using the results of
preceding paragraph \ref{arbi} on the non-tridiagonal sample
Hamiltonian (\ref{dvojaky}) it is easy to verify that during an
exhaustive algebraic solution of Eq.~(\ref{Dieudonnes}) one obtains
the result
 $$\Theta^{(7)}(0,0,0,0,0,0,1)\ \equiv \ {\cal P}^{(7)}\,,$$
i.e.,
one really re-obtains also the initial Hamiltonian-independent ``input''
matrix (\ref{anago}) of parity.

In a more general methodical context it seems tempting to weaken the
constraints imposed upon the pseudometrics. In the PHR framework of
review \cite{ali} precisely this type of generalization (i.e., a
replacement ${\cal P} \to \tilde{\cal P}$ such that not necessarily
$\tilde{\cal P}^2 = I$) was proposed by Mostafazadeh.
%
%
%
For our concrete toy-model $H^{(7)}(R)$ of Eq.~(\ref{dvojaky}), for
example, this would mean that whenever necessary, one could search
for several alternative, multiparametric and Hamiltonian-dependent
sparse-matrix pseudometrics.

 \begin{figure}[h]                     
 \begin{center}                         
 \epsfig{file=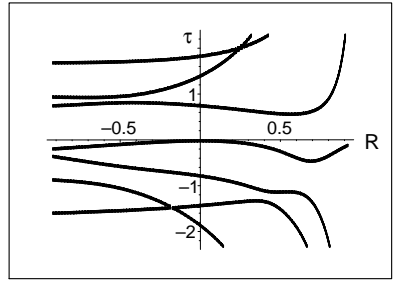,angle=270,width=0.3\textwidth}
  \end{center}                         
 \vspace{-2mm}\caption{The ${R}-$dependence of the
 eigenvalues of the pseudometric candidate (\ref{vejire}).
 \label{fi011}}
 \end{figure}

Obviously, the most natural pseudometric candidate would be  matrix
$\Theta^{(7)}(0,0,0,0,0,1,0)$ which has the compact form
 \be
 \left[ \begin {array}{ccccccc}
 0&0&0&0&0&1&0
 \\\noalign{\medskip}0&0&0
 &0&{\frac {1}{1-R}}&{\frac {R}{R-1}}&1
 \\\noalign{\medskip}0&0
 &0&{\frac {1}{1-R}}&{\frac {R}{R-1}}&{\frac {1}{1-R}}&{\frac {R}{R-1}}
 \\\noalign{\medskip}0&0&{\frac {1}{1-R}}
 &{\frac {R}{R-1}}&{\frac {1}{1-R}}&0&0
 \\\noalign{\medskip}0&
 {\frac {1}{1-R}}&{\frac {R}{R-1}}&{\frac {1}{1-R}}
 &0&0&0
 \\\noalign{\medskip}1&{\frac {R}{R-1}}&{\frac {1}{1-R}}&0&0&0&0
 \\\noalign{\medskip}0&1&{\frac {R}{R-1}}&0&0&0&0\end {array}
 \right]\,.
 \label{vejire}
 \ee
Figure~\ref{fi011} demonstrates that at non-vanishing and not too
large $R\neq 0$ this matrix is invertible and, hence, it could play
the role of an indefinite pseudometric. Still, a serious practical
drawback of the latter pseudometric candidate with eigenvalues
$\tau$ lies in its ill-conditioned invertibility caused by the
smallness of the middle eigenvalue $\tau_3$ in a fairly broad
vicinity of $R=0$. For a remedy we may replace Eq.~(\ref{vejire}),
say, by the following one-parametric family of more general
candidates for the Krein-space pseudometric,
 \ben
 \tilde{\cal P}=\tilde{\cal P}(\xi)
 =\Theta^{(7)}(0,0,0,0,0,1,0)+
 \xi\,{\cal P}\,.
 \een
We could search for a value of $\xi_{optimal}$ which would
(approximatively) maximize the distance of the smallest eigenvalue
of such a generalized pseudometric candidate $\tilde{\cal P}(\xi)$
from zero. In a preparatory step we shall recall the approximate
$R-$independence of $\tau_3$ in Fig.~\ref{fi011} at small $R$ and at
$\xi=0$. This enables us to simplify our optimization problem by
replacing variable $R$ by zero. In other words, a good guess of
$\xi_{optimal}$ will be provided by the value of $\xi_{approximate}$
obtained after the replacement of the true secular polynomial
 $f(\xi,R,\zeta)=\det \left (\tilde{\cal P}(\xi)-\zeta\,I^{(N+1)}
 \right )$
by its reduced two-variable form $f(\xi,0,\tau)$.

A serendipitious benefit of the $R=0$ simplification arises via the
following closed-form factorization formula
 \be
   f(\xi,0,\tau)=
  \left( \tau+\xi \right)  \left(
  \left( \tau+\xi \right)^2
    -2\right)
    \left(
    \left(\left( \tau-\xi \right)^2
    -2\right)^2-2
     \right)\,.
 \ee
This formula enables us to find the recommended value of $\xi$
(which would give {all} eigenvalue roots $\tau$ sufficiently far
from zero) in two alternative and in absolute value equally large
closed forms $\xi_{approximate}=\pm { {1}/{\sqrt {4+\sqrt {8}}}}
 \approx \pm 0.38$
of the point of intersection of the fourth root $\tau_3(\xi)=-\xi$
of $f(\xi,0,\tau)$ with the third root
$\tau_2(\xi)=\xi-\sqrt{2-\sqrt{2}}$ or with the fifth root
$\tau_4(\xi)=\xi+\sqrt{2-\sqrt{2}}$, respectively.

%
%
%

\section{Summary \label{summary}}

In the conclusion one may predict that our present success in
keeping the multi-parametric discrete square well still friendly and
tractable by non-numerical (or, at worst, semi-numerical) means
could have opened a way towards its further future study. Not only
in the obvious mathematical direction towards the continuous-limit
bound-state regime with $N = \infty$ but also towards establishing
its phenomenological relevance. In the dynamical context of
scattering, after all, the first results and an independent
inspiration could be already found in Ref.~\cite{Hernandez}.

Another, independent, implementation-oriented appeal of our
present general
endpoint interactions
may be found in the context of the well known
phenomenological
Su-Schrieffer-Heeger Hamiltonians,
the tridiagonality of which is often strongly violated
by an addition of elements in
the upper-right corner and/or in the lower-left corner of the
Hamiltonian matrix (cf., e.g., Ref.~\cite{SSH}).
Naturally, such an exceptionally remote endpoint
interaction
already acquires a new physical meaning.
Indeed, the main purpose of its absence or presence
gets shifted
from its present action-at-a-distance
interpretation to the
possibility of control
of
the
transition between the
open-chain model of Eq.~(\ref{LuKa8})
and its topologically nonequivalent
closed-loop alternative.

The simplest
sample of the present endpoint interactions
in such a new, spatial-topology-related role would be provided
by
the use of the mere
two extreme elements $b_N\neq 0$ and $d_0\neq 0$ in
interaction matrix Eq.~(\ref{VeKa8}),
or of $a\neq 0$ and $b\neq 0$ in Eq.~(\ref{HaKa7}).
The addition of more (or, perhaps, all) non-vanishing endpoint elements
would lead, in this sense, to a simulation of a toroidal
geometry of the interactions.
An even stronger emphasis put on
the similar simulation of topological effects via
long-ranged {\em alias} endpoint-matching interactions
may be found in the context of quantum graphs \cite{Nizhnik}.
One can summarize that in all of the similar
multiparametric families of mutually related
models the
information about the change of the topology
and/or spectrum is carried
strictly by the
nonlocal boundary conditions.


Needless to add that in many physical applications of the THS
pattern (like, e.g., in the so called interacting boson models of
heavy atomic nuclei \cite{Geyer} or in the descriptions of the
schematic non-Hermitian Lipkin-Meshkov-Glick manybody model
\cite{Geyer,NimrodTony}, etc) the locality/nonlocality is not an
issue at all. After development and description of methodically
oriented nonlocal toy models, one may feel encouraged to replace the
$F-$local external forces $V(\xi)$ by their various $F-$nonlocal,
integral-kernel generalizations $V(\xi,\xi')$ also in some more
realistic toy-model considerations. In the related future work it
will only be necessary that the operators of observables still
remain sufficiently user-friendly.

\newpage

\section*{Appendix A.
Operators of observables in non-Hermitian representations}

According to the conventional model-building strategy the physical
Hilbert space ${\cal H}^{(P)}$ of quantum theory is chosen, {\it a
priori}, in one of its simplest realizations. Typically, one decides
to work with the space $L^2(\mathbb{R})$ of all square-integrable
complex functions of coordinate or momentum. On this background, the
Bender's and Boettcher's discovery \cite{BB} of the possibility of a
peaceful coexistence of observability with non-Hermiticity  changed
our perception of what is an optimal mathematical description of a
quantum system.

The deeper and more detailed account of the amended theory and/or of
its various innovative applications may be found elsewhere
\cite{Carl,ali}.  It has been clarified that a (mathematically more
or less trivial) replacement of the ``primary'' Hilbert space ${\cal
H}^{(P)}$ by its ``secondary'', {unitarily equivalent} alternative
${\cal H}^{(S)}$ might exhibit multiple specific merits even if the
latter physical Hilbert-space requires a nontrivial definition of
the inner product and even if it ceases to belong, therefore, among
the simplest possible realizations of the picture.

In the new theory one has to proceed, basically, in two steps.
Firstly, one has to replace the initial (but, by assumption,
prohibitively complicated) physical Hilbert space ${\cal H}^{(P)}$
by its intermediate substitute ${\cal H}^{(F)}$ which is manifestly
unphysical but, by assumption, {\em much } friendlier. In the
notation as introduced in \cite{SIGMA} this is the first step
towards simplification which is to be achieved by means of a
suitable invertible mapping $\Omega$.

In the second step one takes into account that whenever the
auxiliary map $\Omega$ itself appears non-unitary, the new
representation $H$ of a hypothetical Hamiltonian $\mathfrak{h}$
which was selfadjoint in the prohibitively complicated ${\cal
H}^{(P)}$  becomes non-Hermitian in unphysical ${\cal H}^{(F)}$,
 \be
 H=\Omega^{-1}\mathfrak{h}\Omega \neq \Omega^\dagger
 \mathfrak{h}\left (\Omega^{-1}\right )^\dagger=H^\dagger\,.
 \label{Dysonnes}
 \ee
From the latter relations we may eliminate $\mathfrak{h}$, introduce
an abbreviation $\Omega^\dagger\Omega=\Theta$ and get the original
Hermiticity relation $\mathfrak{h}=\mathfrak{h}^\dagger$ rewritten
in the form of Dieudonn\'{e}'s constraint (\ref{Dieudonnes}). Thus,
in many cases of interest we may define a new operator
$H^\ddagger=\Theta^{-1}H^\dagger\,\Theta$, require that
$H=H^\ddagger$ and declare our new Hamiltonian {\em self-adjoint} in
the third Hilbert space ${\cal H}^{(S)}$ endowed with the {\em ad
hoc} inner-product metric operator $\Theta \neq I$.

The $S-$superscripted space coincides with ${\cal H}^{(F)}$ as a
topological vector space. These two Hilbert spaces merely differ
from each other by the different definitions of the respective
Hermitian conjugations, i.e., of the respective inner products,
 \be
 \left [\br \psi_1|\psi_2\kt \right ]^{(F)} \neq
 \left [\br \psi_1|\psi_2\kt \right ]^{(S)}\ \equiv \
 \left [\br \psi_1|\Theta|\psi_2\kt \right ]^{(F)}\,.
 \ee
In opposite direction, once we fix the metric (with properties
$\Theta=\Theta^\dagger>0$ etc \cite{Geyer}), any candidate $\Lambda$
for an operator of an observable must be essentially self-adjoint in
${\cal H}^{(S)}$. This means that we must demand either that
$\Lambda=\Lambda^\ddagger$ (when considered in the $S-$superscripted
physical Hilbert space), or that
 \be
 \label{otherobs}
 \Lambda^\dagger\Theta=\Theta\,\Lambda
 \ee
when our quantum system is represented in the $F-$superscripted,
calculation-friendly space ${\cal H}^{(F)}$ (which is the preferred
choice, after all, in the vastest majority of applications).

\section*{Appendix B. An explanatory remark on
the concept of locality in non-Hermitian context\label{neloka}}

The current acceptance of the compatibility of non-Hermiticity with
observability was perceivably accelerated by the Bender's and
Boetcher's \cite{BB} restriction of attention to Hamiltonians of
ordinary differential PTS form
 \be
 H^{(BB)}=-\frac{d^2}{d\xi^2}+V^{(BB)}(\xi)\,.
 \label{BBH}
 \ee
These Hamiltonians were defined in the most common Hilbert space
$L^2(\mathbb{R})$, but this space (i.e., in our present notation,
${\cal H}^{(F)}$) had to be reclassified as unphysical. One must
Hermitize $H^{(BB)}$ in ${\cal H}^{(S)}$ by using an amended metric
$\Theta\neq I$.

One of the most important {\em phenomenological\,} consequences of
the acceptance of the PTS theory is that in operators (\ref{BBH}) as
well as in wave functions $\psi(\xi)$, the variable $\xi$ {\em does
not} carry, in general, the habitual meaning of an observable
point-particle coordinate anymore (cf. \cite{171}). The reason lies
in the generic violation of its observability status as given by
condition~(\ref{otherobs}). Indeed, for any operator of position $Q$
defined in ${\cal H}^{(F,S)}$ we may immediately rewrite the latter
condition as follows,
 \be
 Q=\Omega^{-1}\mathfrak{q}\Omega \neq Q^\dagger=\Omega^\dagger
 \mathfrak{q}\left (\Omega^{-1}\right )^\dagger\,.
 \label{beDysonnes}
 \ee
In the next step  we turn attention to space ${\cal H}^{(P)}$ in
which we are allowed to use the principle of correspondence. This
enables us to identify the self-adjoint operator $\mathfrak{q}$ with
an operator of position. In the final step we notice that its
spectrum (i.e., real eigenvalues $q$) coincides with the spectrum of
$Q$ so that up to the trivial cases characterized by the
diagonal-matrix Dyson's maps $\Omega$, the property of locality of
an operator of interaction in ${\cal H}^{(P)}$ will translate into
its nonlocality in ${\cal H}^{(S,F)}$ and {\it vice versa}.

In a constructive manner the latter apparent paradox has been
clarified in Ref.~\cite{Batal} where the authors emphasized that the
concept of a point-particle position {\em alias}  spatial
{coordinate} can only be consistently introduced in the physical
Hilbert space with trivial metric (i.e., in ${\cal H}^{(P)}$). In
opposite direction the illustration of the paradox may be provided
by the most popular Bender's and Boettcher's toy-model (\ref{BBH})
in which the potential $V^{(BB)}(\xi)$ was chosen {local} in
unphysical space ${\cal H}^{(F)}$. Still, the model itself remains
{nonlocal\,} in {\it both} of the physical spaces ${\cal H}^{(P)}$
(where the nonlocality is due to the non-diagonality of $\Omega$ in
Eq.~(\ref{beDysonnes})) and ${\cal H}^{(P)}$ (where the nonlocality
is due to the non-diagonality of product
$\Theta=\Omega^\dagger\Omega$ \cite{171}). We may summarize that in
the model-building context, the locality/nonlocality of
non-Hermitian $V$ in ${\cal H}^{(F)}$ does not change the generic
{nonlocal} nature of the {\em physical\,} lower-case interaction
operator
 \be
 \mathfrak{v}^{(nonlocal)}\ \equiv \
 \Omega\,V^{(local/nonlocal)} \Omega^{-1} \,
 \label{nelo}
 \ee
which acts in ${\cal H}^{(P)}$ and is Hermitian there. Once deduced
from the PTS models, this interaction term is generically nonlocal,
i.e.,  {\em almost always} nonlocal, up to a few very rare
exceptions like the one described in Ref.~\cite{BG}.

Naturally, the generic nonlocality of the action of forces
(\ref{nelo}) is a serious obstacle for their being put in existence
in experimental setups. Even for theorists, such a spatial
nonlocality of action appears counterintuitive \cite{Jones}, mainly
due to the widespread habit of keeping the contact with point
particles (and, hence, of using the principle of correspondence) --
which need not remain a successful strategy even in quantum models
which remain safely Hermitian \cite{Hoo}.

Incidentally, the clear-cut separation of the physical concept of
observable locality (taking place in ${\cal H}^{(P)}$) from its
purely formal parallel in ${\cal H}^{(F)}$ offers also an answer to
the authors of the recent letter \cite{Lee} who claimed,
erroneously, that their results ``essentially kill any hope of PTS
quantum theory as a fundamental theory of nature''. The core of the
misunderstanding clearly lies in the fact that the authors of {\it
loc. cit.} deduced their message from an extremely unfortunate
mix-up of the concepts of locality in ${\cal H}^{(P)}$, in ${\cal
H}^{(S)}$ and in ${\cal H}^{(F)}$ \cite{nelee}.

Let us add that the same inessential role was also played by the
locality of non-Hermitian potentials $V(\xi)$ in the effective,
incomplete-information quantum models of Refs.~\cite{Jones,Cannata}
in which a non-unitarity of the scattering has been admitted. In our
present, more conventional, unitary bound state setting the key
benefit of the above analysis lies in the conclusion that the
locality of potentials $V^{(BB)}(\xi)$ did not have any immediate
observational meaning and it was only relevant due to the {\em
facilitated solvability} of the related Schr\"{o}dinger equation.

\newpage

\end{document}